\documentclass[reprint,prl,amsmath,amssymb,superscriptaddress]{revtex4-2}

\usepackage{graphicx}% Include figure files
\usepackage{dcolumn}% Align table columns on decimal point
\usepackage{bm}% bold math
\usepackage{xcolor}

\def\Z{\mathbb{Z}}

\def\Q{\mathbb{Q}}

\def\R{\mathbb{R}}

\def\P{\mathbb{P}}

\newcommand{\hatp}[1]{#1'}

\newcommand{\ignore}[1]{}

\begin{document}

\preprint{MIT-CTP/5986}
\hfill MIT-CTP/5986

\title{An elliptic approach to  Reid's fantasy}% Force line breaks with \\
%\thanks{A footnote to the article title}%

\author{Lara B. Anderson}
\email{lara.anderson at vt.edu}
\affiliation{Department of Physics, 
Robeson Hall, 0435, 
Virginia Tech,
850 West Campus Drive,
Blacksburg, VA 24061, USA}

\author{James Gray}
\email{jamesgray at vt.edu}
\affiliation{Department of Physics, 
Robeson Hall, 0435, 
Virginia Tech,
850 West Campus Drive,
Blacksburg, VA 24061, USA}

\author{Richard Nally}
\email{rnally at mit.edu}

\affiliation{MIT Center for Theoretical Physics - a Leinweber Institute,  Cambridge, MA 02139, USA}

\author{Washington Taylor}
\email{wati at mit.edu}

\affiliation{MIT Center for Theoretical Physics - a Leinweber Institute,  Cambridge, MA 02139, USA}

\date{\today}% It is always \today, today,
             %  but any date may be explicitly specified

\begin{abstract}
  It is a long-standing problem to prove that the number of  distinct
  topological types of Calabi-Yau threefolds is finite.  A related
  proposition, Reid's fantasy, conjectures that all Calabi-Yau
  threefolds are connected in a single moduli space through extremal transitions. 
Finiteness of topological types 
has been proven for the class
 of elliptic and
  genus one fibered Calabi-Yau threefolds, which recently have been
  shown to constitute the vast majority of known Calabi-Yau
  threefolds; the moduli space of elliptic CY3's is connected.
  In this  letter, we   %report results indicating
  demonstrate
  that all
  non-fibered Calabi-Yau threefolds in two of the largest known
  classes (toric hypersurfaces and complete intersections in products of projective spaces)
 are connected to fibered Calabi-Yau threefolds  through a simple
  class of geometric transitions involving the shrinking of a single
  divisor from a fibered geometry.  This suggests that non-fibered
  Calabi-Yau threefolds are rare  special cases that are reached by
  simplifying fibered Calabi-Yau threefolds, and 
points to a natural path towards proving finiteness and Reid's fantasy for Calabi-Yau threefolds.
%\begin{description}
%\item[Usage]
%Secondary publications and information retrieval purposes.
%\item[Structure]
%You may use the \texttt{description} environment to structure your abstract;
%use the optional argument of the \verb+\item+ command to give the category of each item. 
 %\end{description}
\end{abstract}

%\keywords{Suggested keywords}%Use showkeys class option if keyword
                              %display desired
\maketitle

%\tableofcontents

\section{Introduction}

Calabi-Yau manifolds play a central role in efforts to connect string
theory to observable physics.  The key features of a Calabi-Yau (CY)
threefold $X$,  namely the existence of a  Ricci-flat metric and SU(3)
holonomy,  enable the compactification of 10D string theory to a
space-time $M_{10} = X \times\R^{1, 3}$ in a way that gives a solution
of Einstein's equations preserving some supersymmetry.  (Note that a CY
threefold has three complex dimensions, and six real dimensions).

While Calabi-Yau threefolds have been the subject of intense study by
both physicists and mathematicians over the last four decades,
 some
fundamental questions about these spaces remain unanswered.  In
particular, it is not yet known whether the number of distinct
topological types of Calabi-Yau threefolds is finite or infinite.
Along related lines, Reid's fantasy \cite{reid1987moduli} conjectures that all Calabi-Yau
threefolds (including both smooth CY3s and spaces with some
appropriate class of singularities)
are connected to one another through various geometric transitions
through intermediate singular geometries,
including, for example, the famous conifold transition (see, e.g., \cite{collins2025introductionconifoldtransitions}, and references therein).
It has long been known that complete intersection CY3s in products of projective space \cite{green1988connecting,green1988possible,candelas1990rolling,wang2016connectedness} and hypersurface CY3s in weighted projective space \cite{chiang1995black} are connected by geometric transitions. 

A large class of Calabi-Yau threefolds that can be analyzed more systematically is the set of Calabi-Yau threefolds that admit the structure
of an elliptic or genus one fibration, i.e., where there is a map
$\pi: X \rightarrow B_2$, with $\pi^{-1}(p)  \cong T^2$ for almost all
points $p \in B_2$ (such a fibration of $X$ is elliptic if there is an
appropriate rational  section $\sigma: B_2
\rightarrow X, \pi \sigma = {\rm id}$).  It has been shown that the
number of topological types of elliptic and genus one fibered
Calabi-Yau threefolds is finite   \cite{Grassi, Gross, kawamata1997cone, filipazzi2024boundedness}. 
Elliptically fibered  CY3s  are all connected through extremal transitions, {\it a la} Reid's fantasy.
Furthermore, recent work has shown that almost all known Calabi-Yau
threefolds  are equivalent through simple ``flop'' transitions to an
elliptic CY3 (99.7\% of CICYs \cite{Anderson:2017aux}, 99.99\% of toric hypersurface
CY3s \cite{Huang-WT-fibers}).
  It is thus natural to ask whether the elliptic and genus one
fibered CY3s in fact represent ``typical'' Calabi-Yau threefolds, and
whether there is a simple way in which the small remaining set of
non-fibered CY threefolds can be connected to the fibered threefolds. 

In this  letter, we show that this is indeed the case for two large classes of Calabi-Yau threefold constructions.
Every single one of
the non-fibered CICY and toric hypersurface Calabi-Yau threefolds can
be connected to a fibered CY3 (up to flop equivalance)
 with a single ``move'' corresponding to
adding a single divisor    through an  appropriate class of singular
transitions.    In %\S~\ref{sec:KS}
the next two sections we describe this connectivity in detail for
these two datasets.  Some discussion and
conclusions are contained in the last section.
%\S~\ref{sec:conclusions}.
A more detailed analysis of many of the specific transitions involved
in this picture will   appear in a longer companion paper \cite{long-paper}.
A complete list of transitions for each ostensibly non-fibered CICY and toric hypersurface CY3, along with software to generate and analyze these transitions, is given in \cite{zenodo-archive}.

%Because elliptically fibered CY3s are known to be finite and connected, one natural avenue to attack these questions for general CY3s is to analyze the ways in which fibered CY3s can be deformed to give nonfibered ones; our approach will be to employ the opposite (but equivalent) strategy of classifying the ways in which fibered CY3s can be built from nonfibered ones.

\section{Non-fibered toric hypersurface CY3s}
\label{sec:KS}

The largest set of known Calabi-Yau threefolds that has been
systematically classified are the hypersurfaces in toric varieties
associated with reflexive   4D polytopes.
A complete classification of the 473.8 million reflexive 4D polytopes
was carried out by Kreuzer and Skarke \cite{Kreuzer:2000xy};  the list
of  these polytopes is available online \cite{KS}, and is widely used
in large-scale studies of Calabi-Yau threefolds.  It has been
shown \cite{Huang-WT-fibers, Abbasi:2025lvn} that all but 29,223 of
these polytopes (99.99\%) admit an obvious fibration
  (i.e., contain a  2D reflexive subpolytope).
  %in the form of a two-dimensional reflexive subpolytope encoding the torus fiber
In the present work,  we have analyzed  all the cases without an obvious
fibration, and found that each can be connected to a fibered polytope by a single transition in at least one way.

%each case there is one or more way in %which %a single transition connects to a fibered polytope.

To describe these results in slightly more detail, we briefly review
some of the technical aspects of the toric hypersurface construction
and the analysis of fibrations; more  background details can be found in, e.g.,
\cite{Skarke:1998, Huang-WT-2018-Long}.  A reflexive 4D polytope $\nabla$ is a
lattice polytope, defined as the convex hull of a set of vertices
$v_i\in\Z^{4}$, with the properties that $\nabla$ contains the origin 0
as a (strictly) interior point, and that the dual polytope defined by
$\nabla^{*} = \{ w \in\Q^{4}: \langle w, v_i \rangle \geq -1\}$ is
also a lattice polytope, from which it follows that $\nabla =
(\nabla^{*})^{*}$.  It was shown by Batyrev \cite{Batyrev:1993oya}
that a generic anticanonical hypersurface in any toric variety formed
from a reflexive 4D polytope $\nabla$ along with cone structure
defined by a suitable (FRST) triangulation gives a smooth Calabi-Yau
threefold $X$, and that topological information including the Hodge
numbers of $X$ can be computed from combinatorial information
about the polytope. Similar results hold for more general (``vex'')
triangulations \cite{Berglund:2016nvh,MacFadden:2025ssx}.
Of particular relevance to the analysis here is the Hodge number
$h^{1,1} (X)$, which counts the number of independent divisors 
 (i.e. algebraic hypersurfaces)  in $X$; in the simplest
 cases (``favorable'' polytopes without points interior to facets), this is simply related to the number $n$ of primitive nonzero lattice points in $\nabla$ through $h^{1,1} (X) = n-4$.
The lattice points in  $\nabla^{*}$
can be associated with  the monomials in the
anticanonical hypersurface defining equation.
 Different
triangulations of the polytope $\nabla$ correspond to 
different CY3s associated with distinct flop
phases in a family of CY3s contained in a common extended K\"ahler
cone.  When $\nabla$ contains a 2D reflexive subpolytope $\nabla_2$,
the associated family of CY3s has at least one flop phase with an elliptic or
genus one fibration structure, constructed from a 
(sometimes vex) triangulation of $\nabla$ \cite{Huang-WT-fibers}. 

We are interested here in a broader set $P$ of lattice polytopes
$\nabla$ where both $\nabla$ and 
the integer dual 
polytope $\nabla^{\circ} = \{ w \in\Z^{4}: \langle w, v_i
\rangle \geq -1\} \subseteq \nabla^*$ 
 contain 0  as a
strictly interior point.  This set includes all reflexive polytopes as
well as some non-reflexive polytopes, and was recently considered  in
a different context in
\cite{Taylor:2025gnp}. When $\nabla$ is not reflexive, it is contained properly in the
reflexive polytope $(\nabla^{\circ})^{\circ}$ \footnote{This result is
proven in \cite{skarke1996weight,batyrev2018stringy}.  In the language
of these papers,
$\nabla$ is {\it almost pseudoreflexive}, and
$(\nabla^{\circ})^{\circ}$ is the {\it pseudoreflexive closure} of
$\nabla$, which is always reflexive for polytopes of dimension $d \leq 4$.}. 
  The set $P$ is connected by a set of moves associated with removal
  of a single vertex from any $\nabla\in P$,
 giving a smaller polytope $\check{\nabla}\subset \nabla$
subject to the condition that $\check{\nabla}$ still properly contains
the origin. 
 An anticanonical hypersurface in a non-reflexive $\nabla
\in P$ is still a Calabi-Yau threefold, but may contain
local orbifold
singularities \cite{artebani2016families}
in any star triangulation,  while vex triangulations can give smooth CY3 hypersurfaces; these phases can be related by bistellar flips.
Such
 orbifold
singularities are natural both in the
context of string theory  and mathematically, 
so it makes sense to include such Calabi-Yau threefolds in the physical moduli space.
The transitions between Calabi-Yau threefolds associated
with removal of a vertex from a polytope $\nabla \in P$ correspond to
geometric transitions in which a divisor in the CY3 shrinks to a point
or a curve, or where there is a conifold or trivial transition.
In general, these transitions are  understood from both the
mathematical and physical points of view \cite{wilson1992kahler},
although a variety of different specific local structures arises in
various of these transitions; we describe the general classes of
these transitions below, with one example, and provide 
both a general  local analysis and more compact examples
%number of
%more detailed examples
 in \cite{long-paper}.

%
%and arise in the course of geometric
%transitions such as the well-studied conifold transition.  
%While the exact local structure of these singularities varies,
%and we have not investigated each of these cases in detail, we have
%analyzed
%some general classes and specific examples of the polytopes and transitions
%relevant for this work and have found sensible physical/geometric
%interpretations in all cases we have looked at \cite{long-paper}.
%\wati{ modified the wording about this, which basically applies to the
%$\Delta h^{1,1} >1$ cases, should think about exactly what we want to
%  say here.}

The fact that all polytopes in $P$ are connected through simple
geometric transitions  demonstrates the
connectivity of all toric hypersurface Calabi-Yau threefolds, as long
as we include the singular threefolds associated with intermediate
non-reflexive $\nabla$.  
A related argument suggesting connectivity of
toric hypersurface CY3s was made in \cite{Avram_1996,Kreuzer:2000xy} using subset
relations between reflexive polytopes, but without any analysis
of the more complex singularities that arise in that context this was only a heuristic picture.  The
inclusion of intermediate non-reflexive polytopes associated with
Calabi-Yau threefolds having orbifold singularities, and the detailed
understanding of transitions involving single-ray removal in the context of the intrinsic geometry of Calabi-Yau hypersurfaces rather than just polytopes, however,
provides a precise context for understanding this connectivity both
mathematically and physically.

A central result   of this paper is the following:  for  each of the
29,223 reflexive polytopes $\nabla$ that do not have an obvious fibration (2D
reflexive subpolytope), there is  at least one way in which a single
lattice ray  $r$ can be added, giving a new  (possibly non-reflexive)
polytope $\hatp{\nabla}$
that is fibered.  In terms of Calabi-Yau
threefolds, this means that every  CY3 $X$ in the KS database that does
not admit an obvious fibration can be described in terms of an
elliptic or genus one fibered, but possibly singular, CY3 via a
geometric transition that involves shrinking at most one divisor of
the fibered threefold, %so that $0 \le
                       %h^{1,1}(\hatp{X})-h^{1,1}(X)\le1$,
 possibly along with one or more flops before the divisor is shrunk.
This establishes the primary claim that
 every % flop family of CY3s
CY3
constructed from a non-fibered polytope 
% contains at least one phase/CY3 that
 can be realized by shrinking a single divisor in a
(possibly orbifold-singular) CY3 that is itself
fibered or is flop equivalent to a
fibered CY3.

For each non-fibered $\nabla$, we have systematically constructed and
analyzed all single ray additions that give a fibered $\hatp{\nabla}\in
P$; we find a total of 857,695 such transitions.
%\richard{Added the total number, which I think was not given before.}  
We describe here briefly some statistics and examples of these
constructions.  For each such transition, in addition to the
possibly
non-reflexive % (possibly singular) CY3
 $\hatp{\nabla}$, we have constructed the minimal reflexive polytope
$\bar{\nabla} = (\hatp{\nabla}^{\circ})^{\circ}$ containing $\hatp{\nabla}$, and computed both $\Delta h^{1,1} = h^{1,1} (\bar{\nabla}) - h^{1,1} (\nabla)$ and the difference $\Delta\rho$ between the number of primitive rays in
$\bar{\nabla}$ and $\nabla$.
This is helpful in understanding transitions where
 $\hatp{\nabla}$ is non-reflexive and
$\hatp{X}$  may be
singular, by connecting to an associated 
reflexive
$\bar{\nabla}$
 associated with a smooth $\bar{X}$
where
all singularities have been resolved.
A complete list of the  119,371
transitions with the smallest value of $\Delta h^{1,1}$ for each
$\nabla$, as well as the code used to produce this list, can be found
in the associated data file \cite{zenodo-archive}.
It is worth noting that in all these transitions, the divisor we are
removing from the fibered CY3 plays the role of a section or
multi-section of the elliptic or genus one fibration.

In general, we can consider the local geometry of the transition in
the ambient toric variety associated with  a specific triangulation
of $\nabla$ when a ray is added by considering the
dimension of the cone in the toric fan that
properly
contains the new ray.  We classify a
transition as type A, B, or C if the cone dimension is 4D, 3D, or 2D
respectively.  Further, if we write the new ray as $r = \sum_i
\lambda_i v_i$, where $v_i$ generate the relevant cone
($\lambda_i >0$  always by definition), we distinguish
cases where all coefficients are $\lambda = 1$ (A, B, C) from cases
with other choices of $\lambda$ (A', B'; this cannot occur for a 2D
cone).  Thus, for example, a particularly simple set of transitions
are those of type A, where a $\P^3$ divisor shrinks to a smooth point
in the ambient space.  We have systematically computed the possible
ambient transition types for all 119,371 ray additions with minimal
$\Delta h^{1,1}$; there are a total of 153,979 different such ambient
transition + cone  configurations possible, with multiple options for some ray additions
since in some cases different triangulations provide different local
cone structure for the transition.

Using the toric fan approach \cite{fulton1993introduction}, we can
relate the local geometry of the various transition types in the
ambient space to transitions in the anticanonical hypersurface CY3.
We summarize the results here, further details are given in
\cite{long-paper}.
 Type A and A' transitions, i.e. those where
a divisor shrinks to a point in the ambient space,
 correspond in the intrinsic Calabi-Yau geometry to type  II
 transitions (in the notation of \cite{wilson1992kahler}), where a
 divisor in the CY3  again shrinks
to a point.    Type B transitions, where
 a divisor shrinks to a  curve
in the ambient space, restrict in the CY3 to
type III  transitions from a divisor to a curve. 
These results follow from the algebraic condition that the full
shrunken divisor in the ambient space must be contained in the
hypersurface CY3.
On the other hand,
type C transitions, where shrinking a divisor  leads to a codimension two
surface in the ambient space,
are slightly more subtle.
Generically, these
 correspond %in the cases we have considered
 to conifold transitions in the Calabi-Yau threefold.
The local geometry of the ambient space can be described in terms of a
pair of divisors on a certain toric surface  encoded in the local
toric fan; when the intersection product of these divisors vanishes,
the transition becomes trivial in the hypersurface Calabi-Yau
threefold.

%\wati{ Are we comfortable with the last sentence without
%  qualification, or should we reinstate ``in the cases we have considered''}
%  \richard{It was never clear to me that we absolutely proved this statement, but maybe James' GLSM analysis did?}
%  \richard{So now we have settled this in the affirmative, right?}
%  \james{I believe so - is everyone happy with that?}
%Further geometric details of these transitions will be described
%elsewhere  \cite{long-paper}.  \richard{Moved this sentence ehre from below, think it fits better here but happy to move back.}

 We can analyze the various transitions from non-fibered polytopes to fibered polytopes systematically,  organized by $\Delta h^{1,1}$; note that in the remainder of the paper ``transitions'' always means a transition from a non-fibered polytope  to a fibered polytope.
Of the 29,223 non-reflexive polytopes, there are 3853  that admit
transitions with $\Delta h^{1,1} = 0$.  In these cases, the only
transitions possible are type C transitions, and we have confirmed by
explicit computation that in all these cases the algebraic condition
is satisfied so that the transition is trivial.  Thus, the flop family
of phases constructed from the nonfibered polytope $\nabla$ overlaps
with that of $\hatp{\nabla}$, and these two polytopes describe CY3s in
the same extended K\"ahler  cone.
In 2540  of the 3853 $\Delta h^{1,1} = 0$ cases,  the polytope $\hatp{\nabla}$ is already
reflexive.    In the remainder of the cases, 1--9
 additional primitive
rays must be added to get a reflexive $\bar{\nabla}$   with  $h^{1,1}
(\bar{X}) = h^{1,1} (X), h^{2,1} (\bar{X}) = h^{2,1} (X)$.
In such cases where $\Delta \rho >1$,
%the local analysis of the additional transitions is
%more subtle, and we have not explicitly analyzed all of these cases,
%but we expect that the extended Kahler cone of the
%initial, ostensibly non-fibered CY3 contains at least one fibered
%phase.
 $\hatp{\nabla}$ is non-reflexive, and further transitions are needed.
In general, these are also  type C transitions that do not change the
topology, but for  $\Delta \rho >2$, there can be multiple sequences
of transitions and these may involve intermediate non-reflexive
polytopes with CY3s  in phases with orbifold singularities.  In any
case, however, we expect that independent of the transition path,
 the extended Kahler cone of the
initial, ostensibly non-fibered CY3 contains at least one fibered
phase.
 We have further
confirmed this in some example cases
by  explicitly computing the Hodge numbers and triple
intersection numbers, and
confirming that these are identical for $\nabla$ and  one flop phase of 
$\bar{\nabla}$, while another flop phase of $\bar{\nabla}$ is compatible with the fibration (admits a Kollar divisor \cite{kollar2015deformations, oguiso1993algebraic, wilson1994existence}).
%, although it is computationally infeasible to check all cases.

Of the  remaining 25,370  non-fibered polytopes,  25,337 admit
transitions with $\Delta h^{1,1} = 1$, leaving only 33 polytopes without a transition with $\Delta h^{1,1}\le1$. Of the polytopes admitting a $\Delta(h^{1,1})=1$ transition, 25,315 can be transitioned to a reflexive polytope with a single move, i.e. with $\Delta\rho=1$.  This is thus the
``typical'' situation.
Note that for these typical cases, as well as the 3853 already-fibered
cases, the class $P$  of non-reflexive polytopes does not play any substantial
role  in proving the primary result;
to connect  non-fibered CY3s  to fibered CY3s,
transitions
from reflexive
to non-reflexive polytopes need only be considered
for the remaining 55 cases ($\sim 0.2\%$ of non-fibered
polytopes). 
While the cases that involve multiple transitions via non-reflexive
polytopes are more complicated, we have explicitly analyzed many of
them (see \cite{long-paper} for examples), and the local structure is
essentially the same as described above in terms of type II, III, and
conifold transitions in the CY3, except that orbifold singularities can
slightly complicate the local geometry.

%One useful perspective on many of these transitions comes from the
%F-theory picture \cite{Vafa-F-Theory, Morrison-Vafa-I, Morrison-Vafa-II, Klevers:2014bqa, Abbasi:2025lvn}: similar local changes in the ambient polytope for
%transitions between elliptically fibered CY3s correspond in the
%F-theory picture to various Higgs, tensor, and matter transitions;
%furthermore, the cases where adding a single ray gives a singular CY3
%and further divisors must be added to get a smooth CY3 have a natural
%parallel in the geometry of automatic enhancement in F-theory
%(\cite{Raghuram:2020vxm,Abbasi:2025lvn}).\wati{ would like to work out
%an explicit example, which might prompt modifying this sentence.}
%\richard{Maybe worth taking one of the automatic enhancement examples
%from our paper and deleting rays to get something
%nonfibered...}\james{Maybe do F-theory picture after doing the
%example - which seems to go better with previous stuff?}

As a simple example  of how a non-fibered Calabi-Yau threefold
connects to a fibered Calabi-Yau threefold through a single transition
in the ambient space, consider the standard example of the quintic
hypersurface.  In toric geometry this corresponds to an initial (non-fibered)
polytope $\nabla$  with vertices 
\begin{align}
\{ v_i\}& =\{(1, 0, 0, 0), (0, 1, 0, 0), \\
&\hspace*{0.4in} (0, 0,
1, 0), (0, 0, 0, 1), (-1, -1, -1, -1)\} \nonumber
\label{eq:}
\end{align}
representing the ambient
space $\P^4$.  
The generic anticanonical hypersurface in the corresponding toric
variety ($\P^4$) is the quintic Calabi-Yau threefold, with $h^{1,1} (X) = 1,
h^{2,1} (X) = 101$,
which does not admit a genus one or elliptic fibration.
We can add a single ray $r
= (-1, -1, 0, 0) = v_3+ v_4+ v_5$  to obtain a new reflexive polytope 
$\hatp{\nabla}$ giving rise to  a Calabi-Yau
threefold $\hatp{X}$ with Hodge numbers $h^{1,1} (\hatp{X}) = 2, h^{2,1} (\hatp{X}) =
86$.
$\hatp{\nabla}$ has an obvious 2D reflexive subpolytope spanned by
$v_1, v_2, r$, and, in the notation of \cite{Abbasi:2025lvn}, is a genus one fibered CY3 with fiber $\P^2 = F_1$
and base $\P^2$.   This is
an example of a type B (3D cone, $\lambda_i = 1$) transition in the
ambient space, which corresponds to a type III transition in the
ambient Calabi-Yau.

\section{Non-fibered CICYs}
\label{sec:CICY}

One of the earliest classes of Calabi-Yau threefolds to be constructed
was the set of Complete Intersection Calabi-Yaus (CICYs) \cite{Yau:1986gu,Hubsch:1986ny,Green:1986ck,Candelas:1987kf} defined by a complete intersection of hypersurfaces in a
product of projective spaces. Such a manifold is described by a `configuration matrix' whose columns encode the multidegrees of the polynomials
defining the complete intersection. For each ambient projective factor
$\P^d$, these degrees must add up to  $d +1$  (by the adjunction
formula).   Thus, for example, the quintic can be described as $[\P^4|
  5].$   A CICY is ``obviously'' genus one or elliptic fibered if row and column permutations can be used to write its
configuration matrix in the form
\begin{eqnarray}
X =\left[ \begin{array}{c|cc} {\cal A} & 0& {\cal F} \\ {\cal B} & {\cal M} & \cal{T} \end{array}\right]\,,
\end{eqnarray}
where $\left[{\cal A}|{\cal F}\right]$ is one dimensional and  forms
the torus fiber, while $[{\cal B}|{\cal M}]$
describes the base of the fibration
 and ${\cal T}$ encodes the twist of the fiber over the base \cite{Gray:2014fla}.
 %The number of inequivalent CICY configuration matrices is 7890 
All possible CICY threefolds are described by one of 7890 configuration matrices
\cite{Candelas:1987kf}; it was
found in \cite{Anderson:2017aux} that all but 53 of these are obviously genus one
fibered.  

A simple class of transitions for CICY threefolds
can be described by a ``splitting'' \cite{Candelas:1987kf}, where an additional $\P^n$ factor
is added to the ambient space, and one of the hypersurface conditions
is split into $n +1$ conditions, each linear in the extra factor.
This construction takes a CICY $X$ through a conifold transition to
another CICY $X'$ with, generally, a larger $h^{1,1}$. For example, the quintic threefold can be connected via a conifold/splitting transition to the a genus one fibered manifold defined by
\begin{equation}
X'=\left[\begin{array}{c|ccc} 
\mathbb{P}^4 & 1 & 2 & 2 \\ 
\mathbb{P}^2 & 1 & 1 & 1 \\ 
\end{array} 
\right]
\end{equation}
with $h^{1,1}(X')=2$,$h^{2,1}(X')=58$.
Of the 53
CICYs lacking an obvious fibration, 48 can be connected to fibered
CICY threefolds with $\Delta h^{1,1} = 1$ by splitting on an
additional $\P^1$ or $\P^2$ factor.  This, again, represents a
transition from a non-fibered CY3 to a fibered CY3 by adding a single
divisor.

The remaining 5 non-fibered CICYs (which do not admit any $T^2$-fibration, obvious or non-obvious \cite{Anderson:2017aux}) can be transitioned to fibered manifolds using a straightforward generalization of the methods used for toric hypersurfaces. As in case of toric
hypersurfaces, a single ray can be added to the initial
ambient reflexive polytope. In every case this ray is the sum of a set of generating rays
from one of the component $\P^n$'s in the original ambient space.  In each case, a single $\Delta h^{1,1}=1$ conifold transition of this type
leads to a fibered complete intersection CY3, exactly as in the preceding section.  Details of the 48
splittings and 5 ray transitions are included in a data file in the
accompanying data and software package \cite{zenodo-archive}. 

\section{Conclusions}
\label{sec:conclusions}

We have shown that for two of the largest classes of known Calabi-Yau
threefolds, the small subset of flop families of CY3's that are not ``obviously''
elliptic or genus one fibered in some phase are each either already fibered in a non-obvious manner, or are connected by a single transition to a
(possibly singular)
 Calabi-Yau threefold with one additional divisor and a fibered  flop
 phase in the extended  K\"ahler cone.
%i.e., $\Delta
%h^{1,1} = 1$.
Combined with the observation that almost all known
Calabi-Yau threefolds are in the finite %and connected
set of
topological types admitting (again, up to flops) an elliptic or genus
one fibration, these results seem to  indicate that non-fibered
Calabi-Yau threefolds are somewhat exotic special cases, all of which
can be realized by simplifying fibered Calabi-Yau threefolds through
shrinking a divisor.  

 This understanding of non-fibered Calabi-Yau threefolds in terms of
 reductions of fibered CY3s suggests a natural strategy for proving
the finiteness of topological types of Calabi-Yau threefolds, and for
realizing Reid's fantasy that these threefolds live in a connected
moduli space, with different topological types connected through
specific types of extremal transitions.
In particular, if it can be proven that every non-fibered CY3 can be
reached  by shrinking a divisor from a CY3  that is flop-equivalent to a fibered CY3, and that there are a
finite number of such reductions for any given fibered  (or flop-equivalent) CY3,
this would give a rigorous proof of finiteness
 %(at least up to flop equivalence)
 for topological types
of all Calabi-Yau threefolds. 

Regarding connectivity, on one hand, by expanding the set of Calabi-Yau threefolds
to include certain singular varieties, we have explicitly shown that
all toric hypersurface CY3's are connected through specific types of
extremal transitions; more generally, since it is known that all
elliptic  %and genus one
fibered CY3's live in a connected moduli space,
demonstrating that
this is also true for  genus one fibered CY3s and that
all non-fibered CY3's are connected in the way
described here to fibered CY3's would complete a rigorous proof of
Reid's fantasy.

Physically, demonstrating the finiteness and connectivity of the space
of Calabi-Yau threefolds  would resolve a long-standing puzzle that
is important for understanding  the structure of the string landscape.
More concretely, since the set of fibered Calabi-Yau threefolds admits
a much stronger and more constructive mathematical classification, 
connecting all Calabi-Yau threefolds to the connected and  perhaps
explicitly  constructible set of fibered CY3's provides a potential
avenue for eventually explicitly constructing all
topological types of CY3s.

The results presented here suggest a number of further directions for
development.
%While we have explicitly identified tens of thousands of
%transitions from non-fibered toric hypersurface CY3s to fibered CY3s,
%and have broadly classified these transitions and considered specific
%examples, a complete local classification of these singular
%transitions would be desirable.
We expect that the type of transitions
to fibered geometries found here will naturally extend to the broader
class of all smooth connected CY3s that are complete intersections in
toric varieties, e.g., with fibers like those found in
\cite{Braun:2014qka}, and like the five special cases of CICYs
analyzed here.  More generally, it would be nice to find a more
intrinsic way of describing these transitions that would apply more
generally for CY3s from knowledge of their topological structure,
and/or from a more abstract mathematical perspective.
The local classification we have described here
of these transitions in terms of type II, III
and conifold transitions in the CY3 may provide a useful step towards
such an intrinsic description.
Such a
generalized intrinsic approach would be needed for a full proof of
finiteness and connectivity for all CY3s.  It has been suggested that
classification of non-algebraic CY3s may present the biggest challenge
to Calabi-Yau classification \cite{gross1997primitive}, and it would
be good to understand transitions to fibered CY3s in this context.
It would also be interesting to understand how non-simply connected
and non-K\"ahler Calabi-Yau threefolds can fit into this story.
 The results described in this paper may also be helpful in
 understanding the physics and geometry of
  various Higgs, tensor, and matter transitions in the F-theory
  picture \cite{Vafa-F-Theory, Morrison-Vafa-I, Morrison-Vafa-II,
    Klevers:2014bqa, Abbasi:2025lvn}, where
  in the toric context these transitions correspond to
  similar local changes in the ambient polytope, which will again have
  interpretation in terms of type II, III and conifold transitions.

\begin{acknowledgments}
We would like to thank Paul Oehlmann and Yinan Wang for helpful discussions.
LA, JG, and WT would like to thank the Simons Center for Geometry and
Physics for hospitality during the initiation of this work.  WT would
like to thank the University of Auckland for hospitality during the
completion of this work.  Some computations in this paper were done
using Julia, Python, Sage, Palp, CYTools, the CICY package \footnote{L. B. Anderson, J. Gray, Y.-H. He, S.-J. Lee, and A. Lukas, CICY package, based on methods
described in arXiv:0911.1569, arXiv:0911.0865, arXiv:0805.2875, hep-th/0703249, hep-th/0702210}, and software developed with the help of Claude Code and Codex.
The work of WT was supported  in part by the
DOE under contract \#DE-SC00012567.
The work of RN was supported through a Pappalardo Postdoctoral Fellowship in the MIT Department of Physics. The work of LA and JG was supported in part by the NSF grant PHY-2310588.

\end{acknowledgments}

%\appendix

%\section{Appendixes}

% pasted bbl file here due to issues with arXiv

%apsrev4-2.bst 2019-01-14 (MD) hand-edited version of apsrev4-1.bst
%Control: key (0)
%Control: author (8) initials jnrlst
%Control: editor formatted (1) identically to author
%Control: production of article title (0) allowed
%Control: page (0) single
%Control: year (1) truncated
%Control: production of eprint (0) enabled
%

%\bibliography{references}% Produces the bibliography via BibTeX.

\end{document}